\documentclass[11pt]{article}
\usepackage{paclic32}
\usepackage{times}
\usepackage{latexsym}
\usepackage{amsmath}
\usepackage{amssymb,mathrsfs,epsfig,epstopdf,amsthm}
\usepackage{graphicx}
\usepackage{multirow}
\usepackage{url}

\title{Affect in Tweets Using Experts Model}

\author{Subba Reddy Oota \\
  IIIT-Hyderabad \\
  Hyderabad, India \\
  oota.subba@students.iiit.ac.in \\\And
  Adithya Avvaru \\
  IIIT-Hyderabad \\
  Hyderabad, India \\
  adithya.avvaru@students.iiit.ac.in \\ \AND
  Mounika Marreddy \\
  IIIT-Hyderabad \\
  Hyderabad, India \\
  mounika.marreddy@research.iiit.ac.in \\\And
  Radhika Mamidi \\
  IIIT-Hyderabad \\
  Hyderabad, India \\
  radhika.mamidi@iiit.ac.in \\}

\date{}

\begin{document}
\def \xx {\mathbf{x}}
\def \yy {\mathbf{y}}
\maketitle

\begin{abstract}
Estimating the intensity of emotion has gained significance as modern textual inputs in potential applications like social media, e-retail markets, psychology, advertisements etc., carry a lot of emotions, feelings, expressions along with its meaning. However, the approaches of traditional sentiment analysis primarily focuses on classifying the sentiment in general (positive or negative) or at an aspect level(very positive, low negative, etc.) and cannot exploit the intensity information. Moreover, automatically identifying emotions like anger, fear, joy, sadness, disgust etc., from text introduces challenging scenarios where single tweet may contain multiple emotions with different intensities and some emotions may even co-occur in some of the tweets. In this paper, we propose an architecture, Experts Model, inspired from the standard Mixture of Experts (MoE) model. The key idea here is each expert learns different sets of features from the feature vector which helps in better emotion detection from the tweet. We compared the results of our Experts Model with both baseline results and top five performers of SemEval-2018 Task-1, Affect in Tweets (AIT). The experimental results show that our proposed approach deals with the emotion detection problem and stands at top-5 results.  
\end{abstract}

\section{Introduction}

Sentiment analysis is one of the most famous Natural Language Processing (NLP) tasks. This task was used in social network services~\cite{pang2008opinion,asur2010predicting}, e-retailing, advertising~\cite{qiu2010dasa,jin2007sensitive}, question answering systems~\cite{somasundaran2007qa,stoyanov2005multi}  and many other domains. It focuses on the automatic prediction of polarity or sentiment on tweets or reviews. While most computer science research in this field has focused on strict positive/negative sentiment analysis, the three dominant theories~\cite{marsella2010computational,stelmack1991galen} of emotion agree that humans express or operate with much more nuanced emotion representations. In other words, tweets or reviews, in recent times, include non-standard representations of emotion like emoticons, emojis etc. This task of sentiment analysis became increasingly complex due to an addition of creatively spelt words (for eg, ``gm" for ``good morning", ``hpy" for ``happy" etc,.) and hashtags, particularly in case of tweets.

The current research of Sentiment Analysis is gearing towards evaluating emotion intensity in a text to identify and quantify discrete emotions which can help in above mentioned applications mentioned and many new ones. Here, intensity refers to the degree or quantity of an emotion such as anger, fear, joy, or sadness. For example, consider the three statements ``The product is awesome and delivery is before time", ``It was waste of money and time" and ``This TV is ok ok product at this budget range". The above 3 statements respectively express the level of satisfaction as very happy, very sad and moderately happy. This illustrates the different intensities of happiness of the particular person. Similarly, a person expresses different intensities of other emotions like anger, frustration etc.

\section{Related Work}
In literature, there has been an increasing focus towards building sentiment classification/prediction models through various approaches like rule mining, machine learning or deep learning. A brief overview of the efforts of scientific community towards sentiment related models can be found in ~\cite{pang2008opinion,paltoglou2010online,wilson2004just,liu2012survey}.

Many prior works of emotion detection have always used manual strategies to map emotion category to emotional expression. However, such manual categorization requires an understanding of the emotional content of each expression, which is time-consuming and an arduous task. In~\cite{warriner2013norms}, emotions are projected as points in 3-dimensional space of valence (positiveness-negativeness), arousal (active-passive), and dominance (dominant-submissive). Using this theory, there is a huge effort on creating valence lexicons like MPQA~\cite{wilson2005recognizing}, Norms Lexicon~\cite{warriner2013norms}, NRC Emotion Lexicon~\cite{mohammad2017emotion}, WordNet Affect Lexicon~\cite{baccianella2010sentiwordnet} and many others. However, these lexicon based approaches usually ignore the intensity of emotions and sentiment, which provides important information for fine-grained sentiment analysis. The current research shifts towards automatic emotion classification which has been proposed for many different kinds of text, including tweets~\cite{mohammad2015using,mohammad2017emotion}. 

Existing approaches to analyze intensity are based simply on lexicons, word-embeddings, combinational features and supervised learning. ~\cite{nielsen2011new} introduced lexicon based methods which rely on lexicons to assign the intensity score of each word in the tweet. However, this method did not consider the semantic information from the text. Some supervised methods like deep neural networks were applied to tweet sentiment analysis to predict the polarity~\cite{dos2014deep}.
Although deep learning methods outperform lexicon based methods as shown in~\cite{dos2014deep}, but could not capture the fine-grained property of the sentiment in a text. To capture this fine-grained aspect of a sentiment,~\cite{mohammad2016sentiment} proposed to identify the intensity of emotion in texts. To further expand the scope of emotion analysis, ~\cite{W17-5205,SemEval2018Task1} introduced EmoInt-2017 and SemEval-2018 shared tasks where the top performing teams use deep learning models such as CNN, RNN, LSTMs~\cite{goel2017prayas,koper2017ims} and classifiers like Support Vector Machine or Random Forest ~\cite{duppada2017seernet,koper2017ims}. In the above two tasks, some participants use an ensemble-based approach by simply averaging the outputs of two top performing models~\cite{duppada2017seernet,DBLP:conf/semeval/DuppadaJH18} and the weighted average of predicted outputs of three different deep neural network based models~\cite{goel2017prayas}. 
The subtasks of SemEval-2018 Task-1, AIT~\cite{SemEval2018Task1} are detailed in Section 3.

The structure of the paper is as follows. In section 3, we describe the dataset. Section 4 describes the approach we are using to build the model, while section 5 discusses the approaches of preprocessing and feature extraction. Section 6 presents comparative results of various models along with the analysis of the results. Section 7 presents concluding remarks and future work. 

\section{Dataset Description}
We used the dataset from SemEval-2018 Task 1: AIT \footnote{\url{https://competitions.codalab.org/competitions/17751\#learn_the_details-datasets}} for training our system. 
There is a total of five subtasks: EI-reg (Emotion Intensity regression), EI-oc (Emotion Intensity ordinal classification), V-reg (Valence regression), V-oc (Valence ordinal classification) and E-c (Emotion multi-label classification). Each subtask has three datasets: train, dev, and test. In this paper, we worked on the all five subtasks mentioned above. The dataset details are briefly shown in Table~\ref{dataset}. 

\begin{table}[ht]
\centering
\begin{tabular}{||c|c c c c||} \hline \hline
\textbf{Dataset} & \textbf{Train} & \textbf{Dev} & \textbf{Test} & \textbf{Total} \\ \hline \hline
EI-reg, EI-oc & & & &\\
anger &1701& 388 &1002 & 3091 \\ 
fear &2252& 389 &986 & 3011 \\ 

joy &1616& 290 &1105 & 2905 \\ 
sadness &1533& 397 &975 & 2905 \\ 
V-reg, V-oc &1181& 886 &3259 & 2567 \\  
E-c &6838 &886 &3259 &10953 \\ \hline\hline
\end{tabular}
\caption{SemEval-2018 Task-1 Dataset Details}
\label{dataset}
\end{table}

\section{Approach}
We took inspiration from the Mixture of Experts (MoE)~\cite{jacobs1991adaptive,nowlan1991evaluation} regression and classification models, where each expert tunes to some set of features out of all the features. 
\subsection{MoE Description}
In this subsection, we briefly describe the MoE model to enable the readers to relate our model to MoE architecture. The MoE architecture consists of a number of experts and a gating network. In MoE, there are parameters for each of the expert and a separate set of parameters for gating network. The expert and gate parameters are trained simultaneously using Expectation Maximization~\cite{jordan1994hierarchical} or Gradient Descent Approach~\cite{jordan1995convergence}.

Consider the following regression problem. Let $X = \{\xx^{(n)}\}_{n=1}^{N}$  are $N$ input vectors (samples) and $Y = \{\yy^{(n)}\}_{n=1}^{N}$ are $N$ targets for each input vector. Then, MoE model is described in terms of parameter $\theta = \{\theta_{g},\theta_{e}\}$ where $\theta_{g}$ is set of the gate parameters and $\theta_{e}$ is sets of the expert parameters. Given a sample $\xx$ from among $N$ samples, the total probability of predicting target $\yy$ can be written in terms of the experts as 
\begin{eqnarray}
\label{eq:prob}
\nonumber
P(\yy|\xx, \theta) &=& \sum_{i=1}^{I} P(\yy, \xx|\theta)  \\
\nonumber
               &=& \sum_{i=1}^{I} P(i|\xx, \theta_{g}) P(\yy|i, \xx, \theta_{e}) \\
               &=& \sum_{i=1}^{I} g_{i}(\xx, \theta_{g}) P(\yy|i, \xx, \theta_{e})
\end{eqnarray}
where $I$ is the number of experts, the function $g_{i}(\xx, \theta_{g}) = P(i|\xx, \theta_{g})$ represents the probability of selecting $i^{th}$ expert given $x$ and $P(\yy|i, \xx, \theta_{e})$ represents the probability of $i^{th}$ expert giving $\yy$ on seeing $\xx$.

The MoE training maximizes the log-likelihood of the probability in equation~\ref{eq:prob} to learn the parameters $\theta$~\cite{yuksel2012twenty}.

\begin{figure*}[h]
\begin{center}
\includegraphics[width=0.8\textwidth]{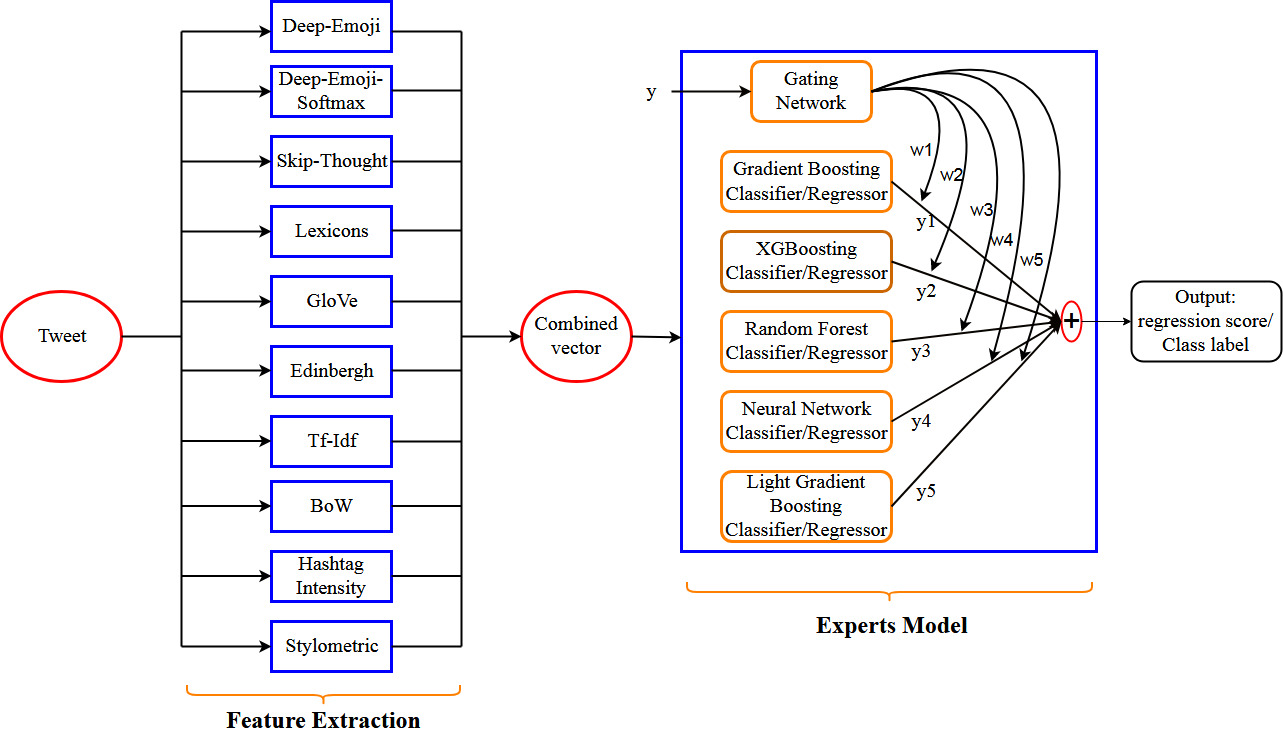}
\caption{ Proposed Experts model}
\label{fig:MOE}
\end{center}
\end{figure*}

\subsection{Our Proposed Approach}
We used the similar architecture, however with some modifications, to measure the intensity of an emotion in a tweet (regression) or predict an emotional intensity (classification). In our proposed approach, we pre-train each expert, to get parameters $\theta_{e}$, on the training samples, unlike traditional MoE model. Each expert, in itself, can be a separate Regression/Classification model like Multi-level Perceptron (MLP) model or Long Short-Term Memory (LSTM) model or any other model that best suits the data and task at hand. Once each expert is trained separately, we train only the gating network based on Gradient Descent Approach. The Detailed description of the model is depicted in Figure~\ref{fig:MOE} and explained below.

We build different models - Neural Network Classifier/Regressor, Gradient Boosting Classifier/Regressor, XGBoost Classifier/Regressor, Random Forest Classifier/Regressor, Lasso Regressor and Light Gradient Boosting Classifier/Regressor and train each of them with the extracted feature vector of each tweet. We obtain this feature vector by concatenating all of the features discussed in Section 5. We assign parameters $\theta_g$, weights and bias, for each Classifier/Regressor at the gating network. Later, we train the gating network to fit the predicted $\hat{y}$ of each expert $i$ with actual $y$ and learn the best $\theta_g$.

Let $w[i]$ denote weight of each expert $i$ at the gating network. Let $b[i]$ denote the bias term for each expert $i$ at the gating network. Let $I$ be the number of experts. We define the Error function ($E$) as \\
\begin{equation*}
E = \sum_{i=1}^{I} \frac{1}{2} prob[i] \Big (\yy[i] - \hat{\yy}[i] + b[i]  \Big )^2
\end{equation*}
where $prob[i]$ is softmax probability of weight $w[i]$, $\yy[i]$ is the actual $\yy$ of $i^{th}$ expert for some sample $\xx$. Similarly, $\hat{\yy}[i]$ is the predicted $\yy$ of $i^{th}$ expert for same sample $\xx$. It is to be noted that $\forall i\hspace{3pt} \yy[i] = \yy$. 

For each sample $\xx$ and $\yy$, we train the gating network using the update equations of gradients as follows:\\
\begin{align*}
\frac{\partial E}{\partial w[i]} &= \frac{1}{2} prob[i] \Big (1 - prob[i] \Big)  \Big(\yy[i] - \hat{\yy}[i] + b[i] \Big)^2 \\
\frac{\partial E}{\partial b[i]} &= prob[i] \Big(\yy[i] - \hat{\yy}[i] + b[i] \Big)
\end{align*}
and
\begin{align*}
w[i] &= w[i] - \eta * \frac{\partial E}{\partial w[i]} \\
b[i] &= b[i] - \eta * \frac{\partial E}{\partial b[i]}
\end{align*}
where $\eta$ is the learning rate.

\section{Preprocessing \& Feature Extraction}
To preprocess each tweet, first we break all the contractions (like ``can't" to ``cannot", ``I'm" to ``I am" etc,.) followed by spelling corrections, decoding special words and acronyms (like ``e g" to ``eg", ``fb" to ``facebook" etc,.) and symbol replacements (like ``\$" to ``dollar", ``=" to ``is equal to" etc). Later, we tokenized each tweet using NLTK tweet tokenizer \footnote{\url{https://www.nltk.org/api/nltk.tokenize.html}}.

The basic idea of using different experts and eclectic features is from the intuition that each expert learn from different aspects of the concatenated features. Hence, we explored and extracted a variety of features; and used only the features which best performs among all the explored ones are explained in the following subsections. 
\subsection{Deep-Emoji Features}
Deep-Emoji~\cite{felbo2017using} performs prediction using the model trained on a dataset of 1246 million tweets and achieves state-of-the-art performance within sentiment, emotion and sarcasm detection. We can use the architecture of Deep-Emoji and train the model using millions of tweets from social media to get a better representation of new data. Using the pre-trained Deep-Emoji model, we extracted two different set of features - one, 64-dimensional vector from the softmax layer and the other, 2304-dimensional vector from attention layer. 

\subsection{Word-Embedding Features}
In this paper, we tried four different pre-trained word-embedding approaches such as Word2Vec~\cite{mikolov2013distributed}, GloVe~\cite{pennington2014glove}, Edinburgh Twitter Corpus~\cite{petrovic2010edinburgh} and FastText~\cite{bojanowski2017enriching} for generating word vectors. We used the GloVe model of 300 dimensions.

\subsection{Skip-Thought Features}
Skip-Thoughts vectors~\cite{kiros2015skip} model is in the framework of encoder-decoder models. Here, an encoder maps words to sentence vector and a decoder is used to generate the surrounding sentences. The main advantage of Skip-Thought vectors is that it can produce highly generic sentence representations from an encoder that share both semantic and syntactic properties of surrounding sentences. Here, we used Skip-Thought vector encoder model to produce a 4800 dimension vector representation of each tweet.

\subsection{Lexicon Features}
We also chose various lexicon features for the model. The lexicon features include AFINN Lexicon~\cite{nielsen2011new} (calculates  positive and  negative  sentiment  scores  from  the  lexicon), MPQA Lexicon~\cite{wilson2005recognizing} (calculates the number of positive and negative words  from  the  lexicon), Bing Liu Lexicon~\cite{bauman2017aspect} (calculates the number of positive and negative words from the lexicon), NRC Affect Intensities, NRC-Word-Affect Emotion Lexicon, NRC Hash-tag Sentiment Lexicon, Sentiment140  Lexicon~\cite{go2009twitter} (calculates  positive  and negative  sentiment  score  provided  by the  lexicon  in  which  tweets  are  annotated by lexicons), and SentiWordNet~\cite{baccianella2010sentiwordnet} (calculates positive, negative, and neutral  sentiment  score). The final feature vector is the concatenation of all the individual features.

\subsection{Hash-tag Intensity Features}
The work by~\cite{mohammad2017emotion} describes that removal of the emotion word hashtags causes the emotional intensity of the tweet to drop. This indicates that emotion word hashtags are not redundant with the rest of the tweet in terms of the overall intensity. Here, we used Depeche mood dictionary~\cite{staiano2014depeche} to get the intensities of hashtag words. We average the intensities of all hashtags of a single tweet to get the total intensity score.

\subsection{Stylometric Features}
Tweets and other electronic messages (e-mails, posts, etc.) are written far shorter, way more informal and much richer in terms of expressive elements like emoticons and aspects at both syntax and structure level, etc. Common techniques use stylometric features~\cite{anchieta2015using} which are categorized into 5 different types: lexical, syntactic, structural, content specific, and idiosyncratic. In this paper, we used 7 stylometric features such as ``number of emoticons", ``number of nouns", ``number of adverbs", ``number of adjectives", ``number of punctuations", ``number of words", and ``average word length".

\subsection{Unsupervised Sentiment Neuron Features}
Unsupervised sentiment neuron model~\cite{radford2017learning} provides an excellent learning representation of sentiment, despite being trained only on the text of Amazon reviews. A linear model using this representation results in good accuracy. This model represents a 4096 feature vector for any given input tweet or text.

\section{Experimental Setup \& Results}

To train our proposed approach, we consider a total of five learning models, one for each expert: Gradient Boosting, XGBoost, Light Gradient Boosting, Random Forest, and Neural Network(NN) for subtasks EI-reg and V-reg.  While for the subtasks EI-oc and V-oc, we consider all the models except NN model. For subtask E-c, we consider all the models except Light Gradient Boosting model.

\begin{table}[ht]
\centering
\begin{tabular}{||c|c||} \hline \hline

\textbf{Model} & \textbf{Parameters}\\ \hline \hline
 & n\_estimators: 3000,   \\ 
Gradient Boosting & Learning rate: 0.05 \\
 & Max\_depth: 4\\ \hline
 & n\_estimators: 100\\  
XGBoosting & learning\_rate: 0.1\\
& max\_depth: 3\\ \hline
& Optimizer: adam \\
Neural Network & Activation : relu\\ \hline
& n\_estimators: 250\\ 
Random Forest & max\_depth: 4 \\ \hline
 & n\_estimators: 720 \\ 
Light Gradient Boosting & learning\_rate: 0.05 \\
& num\_leaves: 5 \\ \hline \hline
\end{tabular}
\caption{Model-Parameters}
\label{model:parameters}
\end{table}

\setlength{\tabcolsep}{1.3pt}

\begin{table*}[ht]
\centering
\begin{tabular}{||c|c c c c c|c c c c c||} \hline \hline

 & \multicolumn{5}{c|}{\textbf{EI-reg (Pearson (all instances))}}    & \multicolumn{5}{c||}{\textbf{EI-reg (Pearson (gold in 0.5-1))}}\\ 
\textbf{Team} & \textbf{macro-avg} & \textbf{anger} & \textbf{fear} & \textbf{joy} & \textbf{sadness} & \textbf{macro-avg} & \textbf{anger} & \textbf{fear} & \textbf{joy} & \textbf{sadness}\\ \hline \hline
SeerNet & 0.799(1) & 0.827 & 0.779 & 0.792 & 0.798 & 0.638(1) & 0.708 & 0.608 & 0.708 & 0.608\\ 
NTUA-SLP & 0.776(2) & 0.782 & 0.758 & 0.771 & 0.792 & 0.610(2) & 0.636 & 0.595 & 0.636 & 0.595\\ 
PlusEmo2Vec & 0.766(3) & 0.811 & 0.728 & 0.773 & 0.753 & 0.579(5) & 0.663 & 0.497 & 0.663 & 0.497\\ 

psyML & 0.765(4) & 0.788 & 0.748 & 0.761 & 0.761 & 0.593(4) & 0.657 & 0.541 & 0.657 & 0.541\\ 
~\textbf{Experts Model} & ~\textbf{0.753(5)} & ~\textbf{0.789} & ~\textbf{0.742} & ~\textbf{0.748} & ~\textbf{0.733} & ~\textbf{0.598(3)} & ~\textbf{0.656} & ~\textbf{0.582} & ~\textbf{0.546} & ~\textbf{0.608}\\ 
Median Team & 0.653(23) & 0.654 & 0.672 & 0.648 & 0.635 & 0.490(23) & 0.526 & 0.497 & 0.420 & 0.517\\ 
Baseline & 0.520(37) & 0.526 & 0.525 & 0.575 & 0.453 & 0.396(37) & 0.455 & 0.302 & 0.476 & 0.350\\  \hline \hline
\multicolumn{9}{c}{Note : The numbers inside parenthesis in both macro-avg columns represent the rank}
\end{tabular}
\caption{Comparison of Regression results of various models with our Experts Model}
\label{EI-reg}
\end{table*}

\begin{table*}[ht]
\centering
\begin{tabular}{||c|c c c c c|c c c c c||} \hline \hline

 &\multicolumn{5}{c|}{\textbf{EI-oc (Pearson (all classes))}}    & \multicolumn{5}{c||}{\textbf{EI-oc (Pearson (some emotion))}}\\ 
\textbf{Team} &\textbf{macro-avg}& \textbf{anger} & \textbf{fear} & \textbf{joy} &\textbf{sadness} & \textbf{macro-avg} & \textbf{anger}& \textbf{fear} & \textbf{joy} & \textbf{sadness}\\ \hline \hline
SeerNet & 0.695(1) & 0.706 & 0.637 & 0.720 & 0.717 & 0.547(1) & 0.559 & 0.458 & 0.610 & 0.560\\ 
PlusEmo2Vec & 0.659(2) & 0.704 & 0.528 & 0.720 & 0.683 & 0.501(4) & 0.548 & 0.320 & 0.604 & 0.533\\ 

psyML & 0.653(3) & 0.670 & 0.588 & 0.686 & 0.667 & 0.505(3) & 0.517 & 0.468 & 0.570 & 0.463\\ 
Amobee & 0.646(4) & 0.667 & 0.536 & 0.705 & 0.673 & 0.480(5) & 0.458 & 0.367 & 0.603 & 0.493\\ 
\textbf{Experts Model} & ~\textbf{0.636(5)} & ~\textbf{0.658} & ~\textbf{0.576} & ~\textbf{0.666} & ~\textbf{0.644} & ~\textbf{0.520(2)} & ~\textbf{0.493} & ~\textbf{0.502} & ~\textbf{0.579} & ~\textbf{0.509}\\ 
Median Team & 0.530(17) & 0.530 & 0.470 & 0.552 & 0.567 & 0.415(17) & 0.408 & 0.310 & 0.494 & 0.448 \\
Baseline & 0.394(26) & 0.382 & 0.355 & 0.469 & 0.370 & 0.296(26) & 0.315 & 0.183 & 0.396 & 0.289\\   \hline \hline
\multicolumn{9}{c}{Note : The numbers inside parenthesis in both macro-avg columns represent the rank}
\end{tabular}
\caption{Comparison of Classification results of various models with our Experts Model}
\label{EI-oc}
\end{table*}

\subsection{Training Strategy}
At the input layer, we used a concatenation vector of all features: Deep-Emoji, Skip-Thought, Lexicons, Stylometric, BoW, Tf-IDF, Glove, Word2Vec, Edinburgh, and HashTagIntensity which is same for each expert. We combined both training and dev data and used them for training our model. The training model is validated by stratified K-fold approach in which the model is repeatedly trained on K-1 folds and the remaining one fold is used for validation. 

In order to tune the hyper-parameters of our experts model, we adopt a grid search cross-validation for each learning model. Using grid search cross-validation, we set the various types of parameters based on the learning model. Table~\ref{model:parameters} shows the parameter settings for all experts.


\subsection{Results}

\begin{table*}[t]
\centering
\begin{minipage}{.45\linewidth}
\centering
\begin{tabular}{||c|c c||} \hline \hline

&\multicolumn{2}{c||}{\textbf{V-reg (Pearson)}} \\
\textbf{Team} &\textbf{(all instances)} & \textbf{(gold in 0.5-1)}\\ \hline \hline
 SeerNet & 0.873 & 0.697 \\ 
 TCS Research & 0.861 &  0.680\\  
PlusEmo2Vec & 0.860 &  0.691\\ 
NTUA-SLP & 0.851 & 0.688\\
 Amobee & 0.843 & 0.644 \\
 \textbf{Experts Model} &\textbf{0.830} & \textbf{0.670}\\
 Median Team & 0.784 & 0.509 \\
 Baseline& 0.585 & 0.449\\  \hline \hline
\end{tabular}
\caption{Comparison of Valence-reg results of various models with our Experts Model}
\label{V-reg}
\end{minipage}
\hspace{0.5cm}
\begin{minipage}{.45\linewidth}
\centering
\begin{tabular}{||c|c c||} \hline \hline

&\multicolumn{2}{c||}{\textbf{V-oc (Pearson)}}\\ 
\textbf{Team} &\textbf{(all instances)} &\textbf{(gold in 0.5-1)}\\ \hline \hline
SeerNet &0.836  &  0.884\\
PlusEmo2Vec & 0.833 &  0.878\\ 
Amobee & 0.813 &  0.865\\ 
psyML & 0.802 &  0.869\\ 
EiTAKA &0.796 & 0.838\\
\textbf{Experts Model} & \textbf{0.738}&\textbf{0.773} \\
Median Team & 0.682 & 0.754\\
Baseline &0.509 &0.560 \\ \hline \hline
\end{tabular}
\caption{Comparison of Valence-oc results of various models with our Experts Model}
\label{V-oc}
\end{minipage}
\end{table*}

\setlength{\tabcolsep}{1.5pt}
\begin{table}[!ht]
\centering
\begin{tabular}{||c|c c c||} \hline \hline

&\multicolumn{3}{c||}{\textbf{E-c}} \\
\textbf{Team} &\textbf{(acc.)} & \textbf{(micro F1)} & \textbf{(macro F1)}\\ \hline \hline
 NTUA-SLP & 0.588(1) & 0.701 & 0.528 \\ 
 TCS Research & 0.582(2) &  0.693 & 0.530\\  
PlusEmo2Vec & 0.576(4) &  0.692 & 0.497\\ 
psyML & 0.574(5) & 0.697 & 0.574\\
 \textbf{Experts Model} &\textbf{0.578(3)} & \textbf{0.691} & \textbf{0.581}\\
 Median Team & 0.471(17) & 0.599 & 0.464 \\
 Baseline& 0.442(21) & 0.570 & 0.443\\  \hline \hline
 \multicolumn{4}{c}{Note : The numbers inside parenthesis in } \\
 \multicolumn{4}{c}{accuracy column represent the rank}
\end{tabular}
\caption{Comparison of E-c results of various models with our Experts Model}
\label{E-c}
\end{table}
 
To evaluate our computational model, we compare our results with SemEval-2018 Task-1 (Affect in Tweets) baseline results, top five performers and Median Team (as per SemEval-2018 results). The results in the EI-reg, EI-oc, V-reg, V-oc, E-c are shown in Tables~\ref{EI-reg},~\ref{EI-oc},~\ref{V-reg},~\ref{V-oc},~\ref{E-c} respectively. The tables illustrate (a) the results obtained by our proposed approach, (b) top five performers in SemEval-2018,  (c) the results obtained by a baseline SVM system using unigrams as features and (d) Median Team among all submissions. From the Tables~\ref{EI-reg} and ~\ref{EI-oc}, we observe that our model (considering only macro-average for Pearson Correlation) for EI-reg and EI-oc stands within 5 places among 48 submissions. A quick walk-through of Table~\ref{EI-reg} for individual emotions shows that anger and fear ranks $3^{rd}$ and $4^{th}$ respectively for EI-reg Pearson(all instances) and for EI-reg Pearson(gold in 0.5-1), fear stands at $3^{rd}$ position and sadness equals score of top performer. Similarly, Table~\ref{EI-oc} for classification results shows that anger and fear ranks $4^{th}$ and $3^{rd}$ respectively for EI-oc Pearson(all classes) and for EI-oc Pearson(some emotion), anger, fear, joy and sadness stands at positions $4^{th}$, $1^{st}$, $4^{th}$ and $3^{rd}$ respectively. 
It is to be noted that in both Tables~\ref{EI-reg} and~\ref{EI-oc}, the numbers inside parenthesis under column ``macro-avg" represent the rank according to macro-avg Pearson scores. These values shows that our model stands at $3^{rd}$ and $2^{nd}$ positions in EI-reg Pearson(gold in 0.5-1) and EI-oc Pearson(some emotion) respectively. 
Tables~\ref{V-reg} and~\ref{V-oc} illustrate that the results from our model are among the top 10 submissions of subtasks V-reg and V-oc. 
Table~\ref{E-c} shows the results of multi-label emotion classification (11 classes). 
Our model is among the top 3 submissions for Jaccard similarity (accuracy) metric, in top 5 for micro F1 metric and topped the submissions for macro F1 metric.

\subsection{Metrics}

\begin{figure*}[!htb]
\centering
\begin{minipage}{.47\textwidth}
\includegraphics[width=\linewidth,height=5cm]{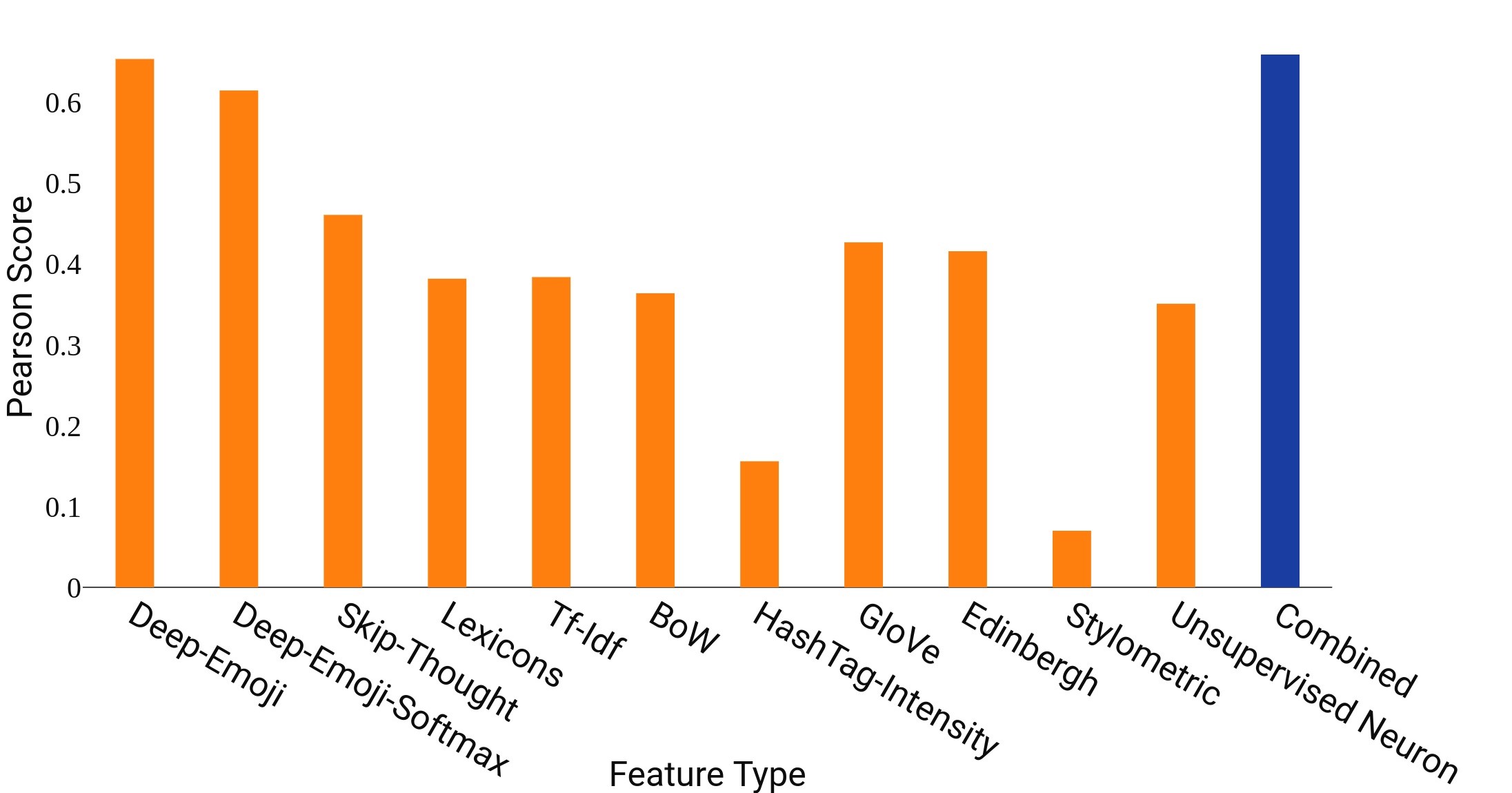}

\begin{center}
(a) Emotion: Anger
\end{center}
\end{minipage}
\qquad
\begin{minipage}{.47\textwidth}
\includegraphics[width=\linewidth,height=5cm]{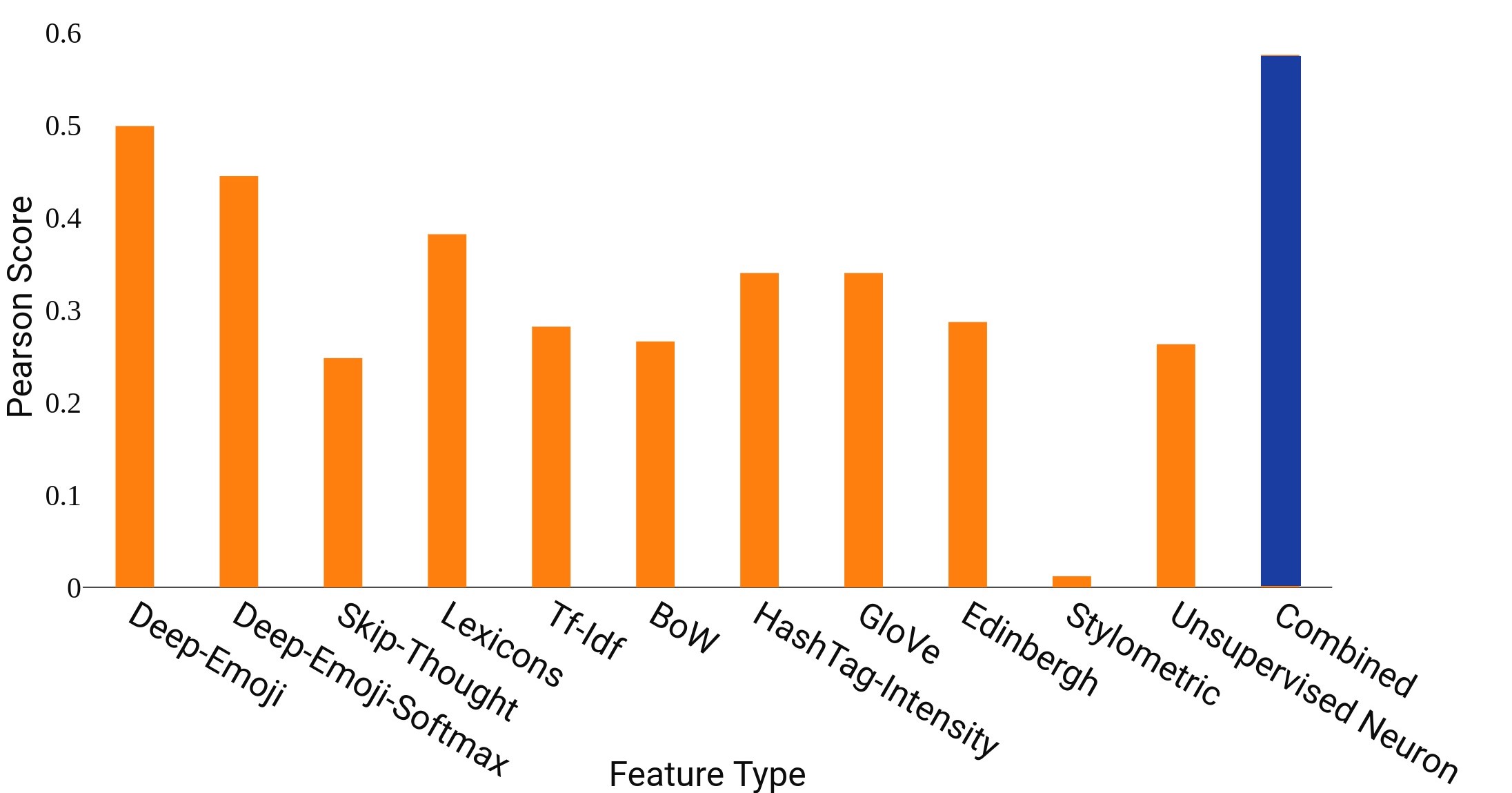}

\begin{center}
(b) Emotion: Fear
\end{center}
\end{minipage}
\qquad
\begin{minipage}{.47\textwidth}
\includegraphics[width=\linewidth,height=5cm]{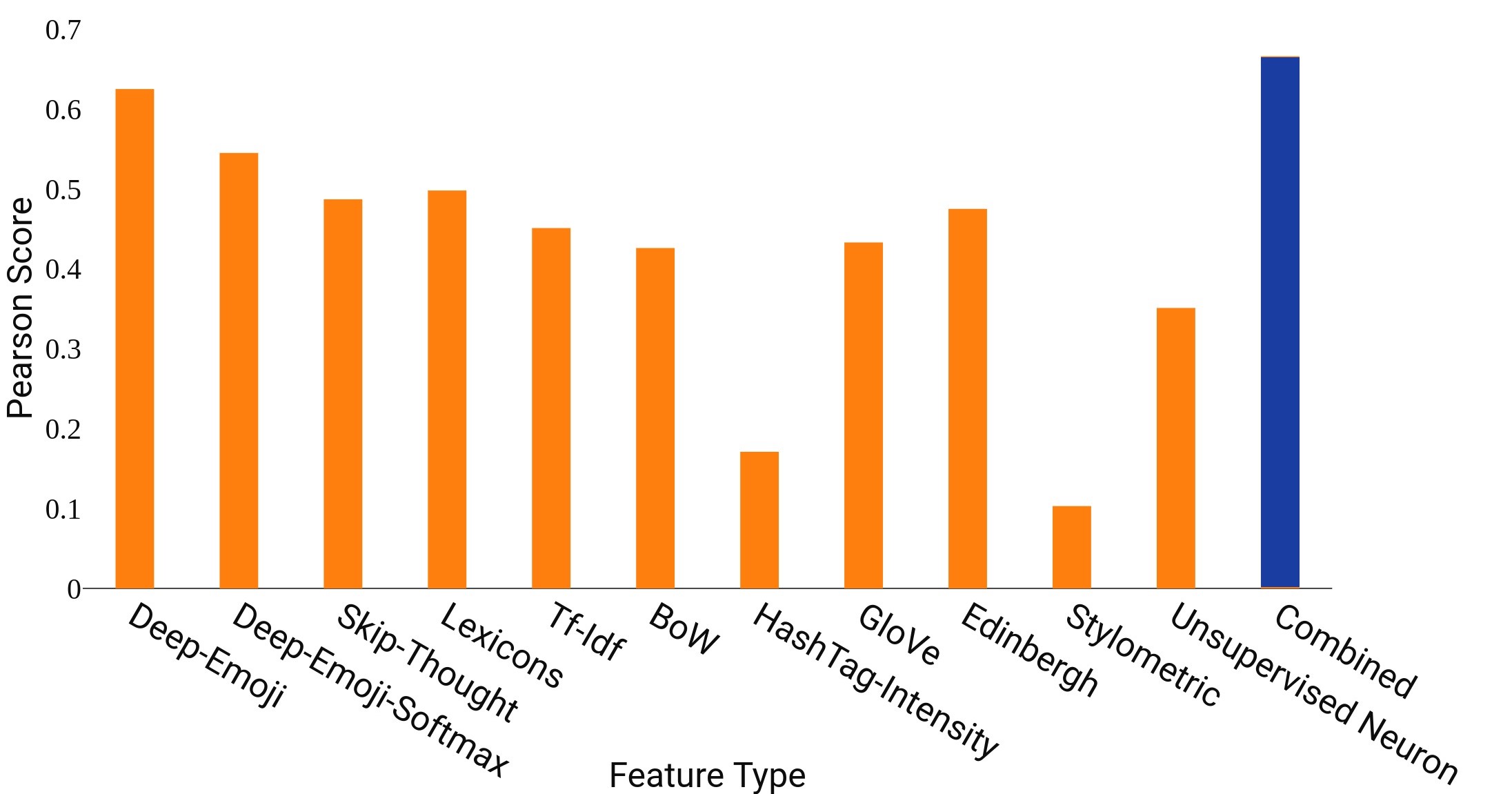}

\begin{center}
(c) Emotion: Joy
\end{center}
\end{minipage}
\qquad
\begin{minipage}{.47\textwidth}
\includegraphics[width=\linewidth,height=5cm]{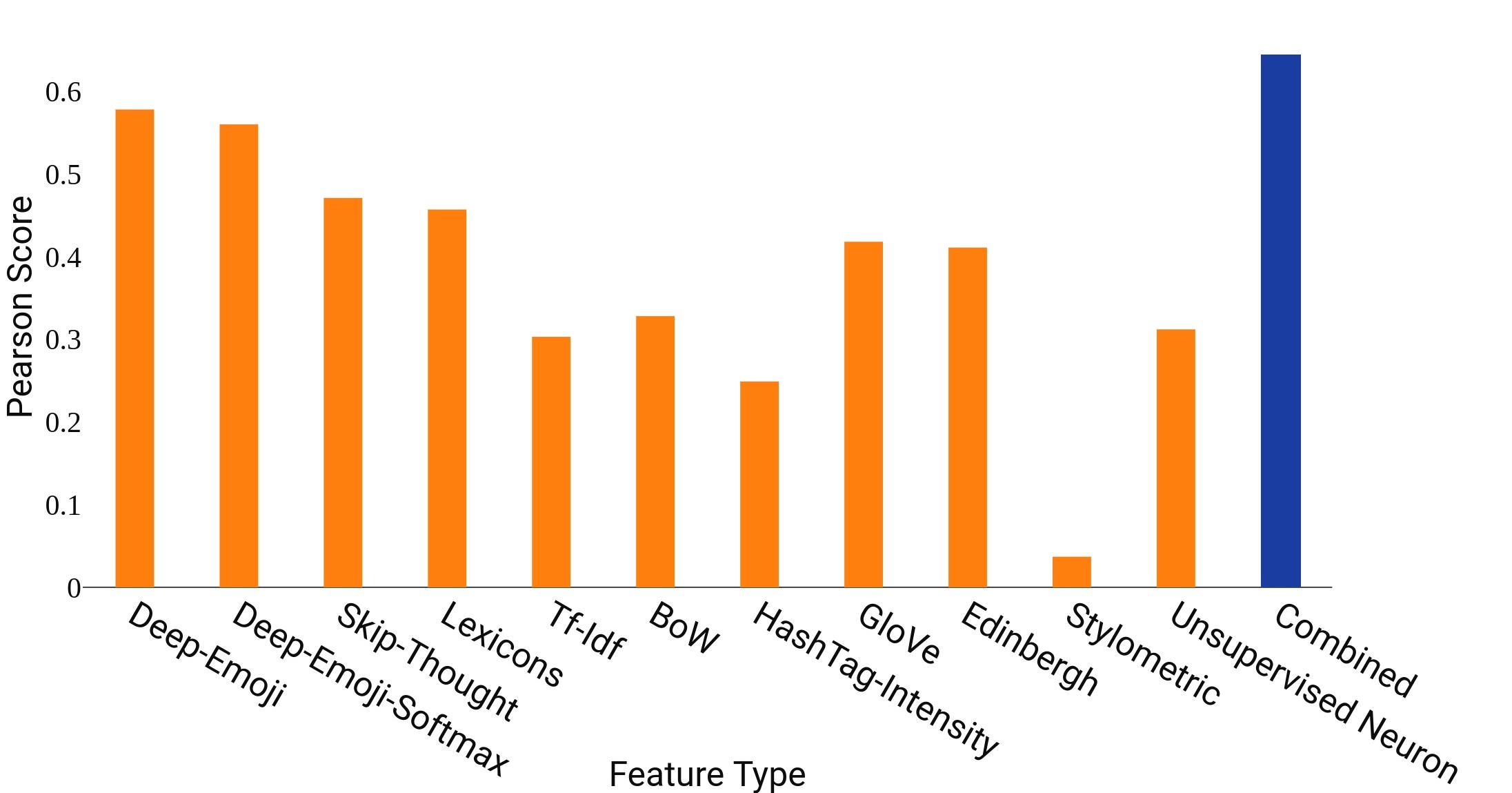}

\begin{center}
(d) Emotion: Sadness
\end{center}
\end{minipage}
\caption{Feature Importance: Comparison of Pearson Scores for each feature vector \& concatenated vector}
\label{fig:2fig}
\end{figure*}

We use the competition metric, Pearson Correlation Coefficient with the Gold ratings/labels from SemEval-2018 task-1 AIT for EI-reg, EI-oc, V-reg and V-oc. Further, macro-average was calculated by averaging the correlation scores of four emotions: anger, fear, joy, and sadness for the tasks EI-reg and EI-oc. Along with Pearson Correlation Coefficient, we use some additional metrics for each subtask. The additional metric used for EI-reg and V-reg tasks was to calculate the Pearson correlation only for a subset of test samples where the intensity score was greater than or equal to 0.5. For the classification subtasks EI-oc and V-oc, we use the additional metric Pearson correlation calculated only for some emotion like low emotion, moderate emotion, or high emotion. However, for the multi label emotion classification E-c, we used the official evaluation metrics Jaccard Similarity (accuracy), micro average F1 score and macro average F1 score of all the classes.

Figure~\ref{fig:2fig} shows the influence of each feature type on scores for predicting the intensity or emotion. We can observe from Figure~\ref{fig:2fig} that for ``Deep-Emoji" and ``Deep-Emoji-Softmax" features, Pearson scores are dominating other feature types. Feature types - Skip-Thought, Lexicons, Glove, and Edinburgh features are contributing approximately similar in each of the 4 emotions. However, Stylometric features and features from Unsupervised sentiment neurons are performing worse.

\section{Conclusion}
In this paper, we have proposed a novel approach inspired from standard Mixture-of-Experts model to predict the intensity of an emotion(Regression) or level of an emotion (Classification) or multi-label emotion classification. Experiment results show that our proposed approach can effectively deal the emotion detection problem and stands at top-5 when compare with SemEval-2018 Task-1 AIT results and baseline results. 
As most of the Pearson scores are in the range of 0.50 to 0.75, there is a lot of scope for improvement in predicting emotions or quantifying the emotion intensity through various other approaches, which are yet to be unfolded. The source code is publicly available at~\url{https://goo.gl/NktJhF} so that researchers
and developers can work on this exciting problem collectively.

\bibliography{paclic32}
\bibliographystyle{acl}

\end{document}